\documentclass[12pt]{article}
\usepackage{amsmath,amssymb,amsthm}
\usepackage{cite}

   \date{\nonumber}
\newcommand{\p}{\partial}
\newcommand{\mb}{\mathbb}
\newcommand{\zt}{\zeta}
\newcommand{\om}{\omega}
\newcommand{\ld}{\lambda}
\newcommand{\ta}{\tau}

\newcommand{\td}{\tilde}
\newcommand{\invp}{D^{-1}}

\newtheorem{prop}{Proposition}
\newtheorem{lemm}{Lemma}

\newtheorem{eg}{Example}

\begin{document}
\title{A new extended KP hierarchy}
\author{ {Xiaojun Liu$^\dag$, Yunbo Zeng$^\ddag$ \thanks{Corresponding author: Y.B. Zeng, Tel:
      +86-10-62787874, Fax: +86-10-62773400. E-mail addresses: lxj98@mails.tsinghua.edu.cn (X.J.  Liu),
      yzeng@math.tsinghua.edu.cn (Y.B. Zeng), rlin@math.tsinghua.edu.cn (R.L. Lin).}, Runliang Lin$^\ddag$
  } \\
  {\small $^\dag$ Department of Applied Mathematics, China Agricultural University,
    Beijing 100083, P.R. China }\\
  {\small $^\ddag$ Department of Mathematical Sciences, Tsinghua University, Beijing 100084, P.R. China} }

\maketitle

\begin{abstract}
  A method is proposed to construct a new extended KP hierarchy, which includes two types of KP equation with
  self-consistent sources and admits reductions to $k$-constrained KP hierarchy and to Gelfand-Dickey
  hierarchy with sources. It provides a general way to construct soliton equations with sources and their Lax
  representations.
\end{abstract}



\section{Introduction}

Generalizations of KP hierarchy attract a lot of interests from both physical and mathematical points of view
\cite{MR638807,MR723457,MR730247,MR688946,MR2006751,MR1621464,MR1629543,MR708435,MR910584,MR940618}. One kind
of generalization is the so-called multi-component KP (mcKP) hierarchy \cite{MR638807}. The mcKP hierarchy
contains many physically relevant nonlinear integrable systems, such as Davey-Stewartson equation,
two-dimensional Toda lattice and three-wave resonant interaction ones. There exist several equivalent
formulations of this mcKP hierarchy: matrix pseudo-differential operator (Sato) formulation, $\ta$-function
approach via matrix Hirota bilinear identities, multi-component free fermion formulation. A coupled KP
hierarchy was generated through the procedure of so-called Pfaffianization \cite{MR1107232}. It was shown in
\cite{MR1738969} that this coupled KP hierarchy can be reformulated as a reduced case of the 2-component KP
hierarchy.  Another kind of generalization of KP equation is the so-called KP equation with self-consistent
sources, which was initiated by V.K.  Mel'nikov \cite{MR708435,MR910584,MR940618}. For example, the first type
of KP equation with self-consistent sources (KPSCS) reads \cite{MR708435,MR910584,MR2077716}
\begin{subequations}
  \label{eqns:KPSCS}
  \begin{align}
    &(4u_{t}-12uu_x-u_{xxx})_x-3u_{yy}+4\sum_{i=1}^N(q_ir_i)_{xx}=0,\label{eqn:KPSCS-main}\\
    &q_{i,y}=q_{i,xx}+2uq_i,\quad i=1,\ldots,N,\\
    &r_{i,y}=-r_{i,xx}-2ur_i.
  \end{align}
\end{subequations}
The second type of KPSCS is \cite{MR708435,Wang-HY}
\begin{subequations}
  \label{eqns:Another-KPSCS}
  \begin{align}
    &4u_t-12uu_{x}-u_{xxx}-3\invp u_{yy}=3\sum_{i=1}^N[q_{i,xx}r_i-q_ir_{i,xx}+(q_ir_i)_y],
    \label{eqn:anoth-KPSCS-main}\\
    &q_{i,t}=q_{i,xxx}+3uq_{i,x}+\frac32q_i\invp u_y+\frac32 q_i\sum_{j=1}^Nq_jr_j+\frac32 u_xq_i,\\
    &r_{i,t}=r_{i,xxx}+3ur_{i,x}-\frac32r_i\invp u_y-\frac32 r_i\sum_{j=1}^Nq_jr_j+\frac32 u_xr_i,
  \end{align}
\end{subequations}
where $\invp$ stands for the inverse of $\frac{d}{dx}$.

The Lax equation of KP hierarchy is given by (see, e.g.,
\cite{MR1964513})
\begin{equation}
  \label{KP-Lax}
  L_{t_n}=[B_n,L],
\end{equation}
where $L=\p+u_1\p^{-1}+u_2\p^{-2}+\cdots$ is a pseudo-differential operator with potential functions $u_i$'s,
$B_n=L^n_+$ stands for the differential part of $L^n$, and $\p=\frac{d}{dx}$. The notation $u_i'=u_{i,x}$ is
used in this paper. The commutativity of $\p_{t_n}$ flows give rise to the zero-curvature equations of KP
hierarchy
\begin{displaymath}
  B_{n,t_k}-B_{k,t_n}+[B_n,B_k]=0.
\end{displaymath}

In this paper, we first introduce a new vector field $\p_{\ta_k}$ which is a linear combination of all vector
fields $\p_{t_n}$. Then we introduce a new Lax type equation which consist of the $\ta_k$-flow and the
evolutions of wave functions. Under the evolutions of wave functions, the commutativity of $\p_{\ta_k}$-flow
and $\p_{t_k}$-flows gives rise to a new extended KP hierarchy. This hierarchy enables us to obtain the first
and second types of KPSCS (i.e., (\ref{eqns:KPSCS}) and (\ref{eqns:Another-KPSCS})) in a different way from
those in \cite{MR708435,MR910584,MR940618,Wang-HY,MR2077716,Xiao-Zeng2005} and to get their Lax
representations directly. This implies that the new extended KP hierarchy obtained in this paper is different from
the mcKP hierarchy given in \cite{MR638807}. Moreover, this new extended KP hierarchy can be reduced to two
integrable hierarchies, i.e., the Gelfand-Dickey hierarchy with self-consistent source (GDHWS)
\cite{MR1165512} and the $k$-constrained KP hierarchy ($k$-KPH) \cite{MR1117170,MR1185854}. The GDHWS includes
the first type of KdV equation with self-consistent sources and the first type of Boussinesq equation with
self-consistent sources. While, the $k$-KPH includes the second type of KdV equation with self-consistent
sources (the Yajima-Oikawa equation) and the second type of Boussinesq equation with self-consistent
sources. Thus, the method proposed in this paper to construct the new extended KP hierarchy provides a general way to
find soliton equation with self-consistent sources as well as their Lax representations. Our paper will be
organized as follows. In section 2, we construct the new extended KP hierarchy and show that it contains the first
and second types of KPSCS. In section 3, the new extended KP hierarchy is reduced to the Gelfand-Dickey hierarchy
with self-consistent source and the $k$-constrained KP hierarchy. In section 4, some conclusions are given.

\section{New extended KP hierarchy}
With the help of formal dressing method, the $L$ operator for KP hierarchy can be written in the following
form \cite{MR1964513}
\begin{equation}
  \label{eqn:dress}
  L=\phi\p\phi^{-1}, \quad \phi=1+w_1\p^{-1}+w_2\p^{-2}+\cdots.
\end{equation}
And the evolution of the dressing operator $\phi$ is given by
\begin{equation}
  \label{eqn:ev-phi}
  \phi_{t_n}=-L^n_-\phi.
\end{equation}
The wave function and the adjoint one are then given by
\begin{displaymath}
  \om(t,z)=\phi\exp\left(\xi(t,z)\right),\quad
  \om^*(t,z)=(\phi^*)^{-1}\exp(-\xi(t,z)),
\end{displaymath}
where $\xi(t,z)=\sum_{i>0} t_iz^i$. They satisfy the following
equations
\begin{subequations}
\label{eqns:ev-wave}
\begin{align}
  Lw(t,z)&=zw(t,z),\quad \frac \p {\p t_n}w(t,z)=B_n(w(t,z)),  \label{eqn:ev-wv}\\
  L^*w^*(t,z)&=zw^*(t,z),\quad \frac \p {\p t_n}w^*(t,z)=-B_n^*(w^*(t,z)).\label{eqn:ev-adj-wv}
\end{align}
\end{subequations}
It was proved in \cite{MR1964513} that the principle part of the
resolvent defined by
\begin{subequations}
\label{eqns:res-lema}
\begin{equation}
  T(z)_-=\sum_{i\in\mb{Z}}L^i_-z^{-i-1},
\end{equation}
can be written as
\begin{equation}
  \label{eq:lema}
  T(z)_-=w(t,z)\p^{-1}w^*(t,z).
\end{equation}
\end{subequations}

For any fixed $k\in\mb{N}$, we define a new variable $\ta_k$ whose vector field
is $$\p_{\ta_k}=\p_{t_k}-\sum_{i=1}^N\sum_{s\ge0}\zt_i^{-s-1}\p_{t_s},$$ where $\zt_i$'s are arbitrary
distinct non-zero parameters. The $\ta_k$-flow is given by
\begin{align*}
  L_{\ta_k}&=\p_{t_k}L-\sum_{i=1}^N\sum_{s\ge1}\zt_i^{-s-1}\p_{t_s}L=[B_k,L]-\sum_{i=1}^N\sum_{s\ge0}\zt_i^{-s-1}[B_s,L]\\
  &=[B_k,L]+\sum_{i=1}^N\sum_{s\in\mb{N}}\zt_i^{-s-1}[L^s_-,L].
\end{align*}
Define $\td{B}_k$ by
\begin{displaymath}
  \td{B}_k=B_k+\sum_{i=1}^N\sum_{s\in\mb{Z}}\zt_i^{-s-1}L^s_-,
\end{displaymath}
which, according to (\ref{eqns:res-lema}), can be written as
\begin{displaymath}
  \td{B}_k=B_k+\sum_{i=1}^Nw(t,\zt_i)\p^{-1}w^*(t,\zt_i).
\end{displaymath}
By setting $q_i=w(t,\zt_i)$, $r_i=w^*(t,\zt_i)$, we have
\begin{subequations}
\begin{equation}
  \td{B}_k=B_k+\sum_{i=1}^Nq_i\p^{-1}r_i,
\end{equation}
where $q_i$ and $r_i$ satisfy the following equations
\begin{equation}
  q_{i,t_n}=B_n(q_i),\quad r_{i,t_n}=-B^*_n(r_i)\quad i=1,\cdots,N.
\end{equation}
\end{subequations}
Now we introduce a new Lax type equation given by
\begin{subequations}
\label{eqns:nLax}
\begin{equation}
  \label{eqn:nLax}
  L_{\ta_k}=[B_k+\sum_{i=1}^Nq_i\p^{-1}r_i,L].
\end{equation}
with
\begin{equation}
  \label{eqns:q-r}
  q_{i,t_n}=B_n(q_i),\quad r_{i,t_n}=-B_n^*(r_i)\quad i=1,\cdots,N.
\end{equation}
\end{subequations}
We have the following lemma \cite{MR1345081}.
\begin{lemm}
  \label{lemm} $[B_n,q\p^{-1}r]_-=B_n(q)\p^{-1}r-q\p^{-1}B_n^*(r)$.
\end{lemm}
\begin{proof}
  Without loss of generality, we consider a monomial: $P=a\p^n$ ($n\ge0$). Then
  \begin{displaymath}
    [P,q\p^{-1}r]_-=aq^{(n)}\p^{-1}r- (q\p^{-1}r a\p^n)_-.
  \end{displaymath}
  Notice that the second term can be rewritten in the following way
  \begin{align*}
    &(q\p^{-1}ra\p^{n})_-=(q\p^{-1}\p(ra)\p^{n-1}-q\p^{-1}(ra)'\p^{n-1})_-\\
    &=(-q\p^{-1}(ra)'\p^{n-1})_-=\cdots=(-1)^nq\p^{-1}(ar)^{(n)}=q\p^{-1}P^*(r),
  \end{align*}
  then the lemma is proved.
\end{proof}
\begin{prop}
  (\ref{KP-Lax}) and (\ref{eqns:nLax}) give rise to the following new extended KP hierarchy
  \begin{subequations}
    \label{eqns:nmcKP}
    \begin{align}
      &B_{n,\ta_k}-(B_k+\sum_{i=1}^Nq_i\p^{-1}r_i)_{t_n}+[B_n,B_k+\sum_{i=1}^Nq_i\p^{-1}r_i]=0\label{eqn:nmcKP-zc}\\
      &q_{i,t_n}=B_n(q_i),\label{eqn:nmcKP-q}\\
      &r_{i,t_n}=-B_n^*(r_i),\quad i=1,\cdots,N.\label{eqn:nmcKP-r}
    \end{align}
  \end{subequations}
\end{prop}
\begin{proof}
  We will show that under (\ref{eqns:q-r}), (\ref{KP-Lax}) and (\ref{eqn:nLax}) give rise to
  (\ref{eqn:nmcKP-zc}). For convenience, we assume $N=1$, and denote $q_1$ and $r_1$ by $q$ and $r$,
  respectively. By (\ref{KP-Lax}), (\ref{eqns:nLax}) and Lemma~\ref{lemm}, we have
  \begin{align*}
    &B_{n,\ta_k}=(L^n_{\ta_k})_+=[B_k+q\p^{-1}r,L^n]_+=[B_k+q\p^{-1}r,L^n_+]_++[B_k+q\p^{-1}r,L^n_-]_+\\
    &=[B_k+q\p^{-1}r,L^n_+]-[B_k+q\p^{-1}r,L^n_+]_-+[B_k,L^n_-]_+\\
    &=[B_k+q\p^{-1}r,B_n]-[q\p^{-1}r,B_n]_-+[B_n,L^k]_+\\
    &=[B_k+q\p^{-1}r,B_n]+B_n(q)\p^{-1}r-q\p^{-1}B_n^*(r)+B_{k,t_n}\\
    &=[B_k+q\p^{-1}r,B_n]+(B_k+q\p^{-1}r)_{t_n}.
  \end{align*}
\end{proof}
Under (\ref{eqn:nmcKP-q}) and (\ref{eqn:nmcKP-r}), the Lax representation for (\ref{eqn:nmcKP-zc}) is given by
\begin{subequations}
\begin{align}
  \psi_{\ta_k}&=(B_k+\sum_{i=1}^Nq_i\p^{-1}r_i)(\psi),\label{eqn:Lax-1}\\
  \psi_{t_n}&=B_n(\psi).\label{eqn:Lax-2}
\end{align}
\end{subequations}
Now, we list some examples in the new extended KP hierarchy
(\ref{eqns:nmcKP}).
\begin{eg}[The first type of KPSCS]
  For $n=2$ and $k=3$, (\ref{eqns:nmcKP}) yields
  \begin{subequations}
    \label{eqns:KPSCS-1}
    \begin{align}
      &u_{1,y}-u_1''-2u_2'=0,\label{eqn:KPSCS-1-1}\\
      &2u_{1,t}-3(u_2+u_1')_{t_2}+3u_2''+u_1'''-6u_1u_1'+2\sum_{i=1}^N(q_ir_i)'=0,\label{eqn:KPSCS-1-2}\\
      &q_{i,t_2}=q_i''+2u_1q_i,\quad r_{i,t_2}=-r_i''-2u_1r_i.
    \end{align}
  \end{subequations}
  Set $y:=t_2$, $t:=\tau_3$, $u:=u_1$, and eliminate $u_2$ by differentiating the second equation with respect
  to $x$, we get the first type of KP equation with self-consistent sources (\ref{eqns:KPSCS}). The Lax
  representation of (\ref{eqn:KPSCS-main}) is
  \begin{align*}
    \psi_y&=(\p^2+2u)(\psi),\\ \psi_t&=(\p^3+3u\p+(\frac32 \invp u_y+\frac32 u_x)+\sum_{i=1}^N q_i\p^{-1}r_i)(\psi).
  \end{align*}
\end{eg}
\begin{eg}[The second type of KPSCS]
  For $n=3$ and $k=2$, (\ref{eqns:nmcKP}) yields
  \begin{subequations}
    \begin{align}
      &u_{1,y}+\sum_{i=1}^N(q_ir_i)'-u_1''-2u_2'=0,\label{eqn:KPSCS-2-1}\\
      &3(u_2+u_1')_y-2u_{1,t}-u_1'''+3\sum_{i=1}^N(q_i'r_i)'+6u_1u_1'-3u_2''=0,\label{eqn:KPSCS-2-2}\\
      &q_{i,t_3}=q_{i,xxx}+3u_1q_{i,x}+(3u_2+3u_1')q_i, \label{eqn:KPSCS-2-q}\\
      &r_{i,t_3}=r_{i,xxx}+3u_1r_{i,x}-3u_2r_i.\label{eqn:KPSCS-2-r}
    \end{align}
  \end{subequations}
  Let $y:=\ta_2$, $t=t_3$, $u:=u_1$, and eliminate $u_2$ by integrating the first equation with respect to
  $x$, we get the second type of KPSCS (\ref{eqns:Another-KPSCS}). This equation was introduced in
  \cite{MR708435}, and rediscovered by source generating method \cite{Wang-HY}. Under (\ref{eqn:KPSCS-2-q})
  and (\ref{eqn:KPSCS-2-r}), the Lax representation for (\ref{eqn:anoth-KPSCS-main}) is
  \begin{align*}
    \psi_y&=(\p^2+2u+\sum_{i=1}^N q_i\p^{-1}r_i)(\psi),\\
    \psi_t&=(\p^3+3u\p+(\frac32\invp u_y+\frac32 u_x +\frac32\sum_{i=1}^Nq_ir_i))(\psi).
  \end{align*}
\end{eg}
\begin{eg}
  For $n=4$, $k=2$, and $N=1$, (\ref{eqns:nmcKP}) gives higher order equations
  \begin{align*}
    &u_{1,y}-u_{1,xx}-2u_{2,x}+(qr)'=0\\
    &2u_{2,y}+3u_{1,xy}+3(qr)''-2(qr')'+8u_1u_{1,x}-3u_{1,xxx}-4u_{3,x}-8u_{2,xx}=0\\
    &2u_{3,y}+3u_{2,xy}+2u_{1,xxy}-u_{1,t}+2(qr)'''-3(qr')''+2(qr'')'+4u_1u_{1,xx}\\
    &+4u_1(qr)'+4u_{1,x}u_2+6(u_{1,x})^2-2u_{3,xx}-3u_{2,xxx}-u_{1,xxxx}=0\\
    &q_t=(\p^4+4u_1\p^2+(4u_2+6u_1')\p+(4u_3+6u_2'+4u_1''+6u_1^2))q\\ 
    &r_t=-(\p^4+4u_1\p^2+(4u_2+6u_1')\p+(4u_3+6u_2'+4u_1''))^*r.
  \end{align*}
  Here $y:=\tau_2$, $t:=t_4$. By using
  \begin{align*}
    2u_2&=\invp u_{1,y}-u_{1,x}+qr\\
    4u_{3,x}&=\invp u_{1,yy}-2u_{1,xy}+ (qr)_y+4(qr)_{xx}-2(qr_x)_x+8u_1u_{1,x}+u_{1,xxx}
  \end{align*}
  the equations yield
  \begin{align*}
    &\frac12\invp u_{1,yyy}+2(u_1^2)_{xy}+\frac12u_{1,xxxy}+\frac32u_{1,xy}-u_{1,xt}\\
    &+6(u_{1,x}^2)_x+4u_1u_{1,xxx}+2u_{1,x}u_{1,y}+2u_{1,xx}\invp
    u_{1,y}-\frac32u_{1,xyy}-2(u_1^2)_{xxx}\\
    &+\frac12(qr)_{yy}+3(qr)_{xxy}-(qr_x)_{xy}-2(q_xr_x)_{xx}\\
    &+6u_{1,x}(qr)_x+4u_1(qr)_{xx}+2u_{1,xx}qr-\frac32(qr)_{xxx}=0, \\
    &q_t=(\p^4+4u_1\p^2+(4u_2+6u_1')\p+(4u_3+6u_2'+4u_1''))q\\
    &r_t=-(\p^4+4u_1\p^2+(4u_2+6u_1')\p+(4u_3+6u_2'+4u_1''))^*r.
  \end{align*}
\end{eg}

\section{Reductions}
The new extended KP hierarchy (\ref{eqns:nmcKP}) admits reductions to several well-known $(1+1)$-dimensional systems.

\subsection{The $n$-reduction of (\ref{eqns:nmcKP})}
The $n$-reduction is given by
\begin{equation}
  \label{eqn:n-redu}
  L^n=B_n \quad \text{or}\quad L^n_-=0,
\end{equation}
then (\ref{eqns:ev-wave}) implies that
\begin{subequations}
  \label{eqns:ev-wave-redu}
  \begin{align}
    &B_n(q_i)=L^nq_i=\zt_i^nq_i,\\
    &-B_n^*(r_i)=-L^{n*}r_i=-\zt_i^nr_i.
  \end{align}
\end{subequations}
By using Lemma 1 and (\ref{eqns:ev-wave-redu}), we can see that the
constraint (\ref{eqn:n-redu}) is invariant under the $\ta_k$ flow
\begin{align}
  &(L^n_-)_{\ta_k}=[B_k,L^n]_-+\sum_{i=1}^N[q_i\p^{-1}r_i,L^n]_-\notag\\
  =&[B_k,L^n_-]_-+\sum_{i=1}^N[q_i\p^{-1}r_i,L^n_+]_-+\sum_{i=1}^N[q_i\p^{-1}r_i,L^n_-]_-\notag\\
  =&\sum_{i=1}^N [q_i\p^{-1}r_i,B_n]_-=-\sum_{i=1}^N (q_{i,t_n}\p^{-1}r_i+q_i\p^{-1}r_{i,t_n})\\
  =&-\sum_{i=1}^N(\zt_i^nq_i\p^{-1}r_i-\zt_i^nq_i\p^{-1}r_i)=0.\label{eqn:n-invar}
\end{align}
The equations (\ref{eqn:n-redu}) and (\ref{eqn:ev-phi}) imply that
$\phi_{t_n}=0$, so $(L^k)_{t_n}=0$, which together with
(\ref{eqn:n-invar}) means that one can drop $t_n$ dependency from
(\ref{eqns:nmcKP}) and obtain
\begin{subequations}
  \label{eqs:GDHSCS}
  \begin{align}
    B_{n,\ta_k}&=[(B_n)^{\frac k n}_++\sum_{i=1}^N q_i\p^{-1}r_i,B_n],\\
    B_n(q_i)&=\zt_i^nq_i,\\
    B_n^*(r_i)&=-\zt_i^nr_i,\quad i=1,\cdots,N.
  \end{align}
\end{subequations}
The system (\ref{eqs:GDHSCS}) is the so-called Gelfand-Dickey
hierarchy with self-consistent sources~\cite{MR1165512}.

For $n=2$ and $k=3$, (\ref{eqs:GDHSCS}) presents the first type of
KdV equation with self-consistent sources ($t:=\ta_3$, $u:=u_1$)
\begin{align*}
  &u_t-3uu_x-\frac14 u_{xxx}+\sum_{i=1}^N (q_ir_i)_x=0,\\
  &q_{i,xx}+2uq_i=\zt_i^2q_i,\\
  &r_{i,xx}+2ur_i=\zt_i^2r_i,\quad i=1,\cdots,N, 
\end{align*}
with Lax representation
\begin{displaymath}
  (\p^2+2u)(\psi)=\ld\psi,\quad \psi_t=(\p^3+3u\p+\frac32u'+\sum_{i=1}^N q_i\p^{-1}r_i)(\psi).
\end{displaymath}
The first type of KdV equation with self-consistent sources can be
solved by the inverse scattering method \cite{MR940618,Lin-2001} and
by the Darboux transformation (see \cite{Lin-2006} and the
references therein).

For $n=3$ and $k=2$, (\ref{eqs:GDHSCS}) presents the first type of
Boussinesq equation with self-consistent sources ($t:=\ta_2$,
$u:=u_1$)
\begin{align*}
  &u_{tt}+\frac13u_{xxxx}+2(u^2)_{xx}+\sum_{i=1}^N(q_{i,x}r_i-q_ir_{i,x})_{xx}+\sum_{i=1}^N(q_ir_i)_{xt}=0,\\
  &q_{i,xxx}+3uq_{i,x}+q_i(\frac32\invp u_y+\frac32 u_x+\frac32\sum_{j=1}^Nq_jr_j)=\zt_i^3q_i,\\
  &r_{i,xxx}+3ur_{i,x}-r_i(\frac32\invp u_y-\frac32 u_x+\frac32\sum_{j=1}^Nq_jr_j)=\zt_i^3r_i,\quad
  i=1,\cdots,N,
\end{align*}
with Lax representation
\begin{displaymath}
  (\p^3+3u_1\p+3u_2+3u_{1,x})(\psi)=\ld\psi,\quad
  \psi_t=(\p^2+2u_1+\sum_{i=1}^N q_i\p^{-1}r_i)(\psi).
\end{displaymath}

\subsection{The $k$-constrained hierarchy of (\ref{eqns:nmcKP})}
The $k$-constraint is given by \cite{MR1345081,MR1117170,MR1185854}.
$$L^k=B_k+\sum_{i=1}^N q_i\p^{-1}r_i$$
By dropping $\ta_k$ dependency from (\ref{eqns:nmcKP}), we get
\begin{subequations}
  \label{eqs:k-constrained}
  \begin{align}
    &\left(B_k+\sum_{i=1}^N q_i\p^{-1}r_i\right)_{t_n}
    =\left[(B_k+\sum_{i=1}^N q_i\p^{-1}r_i)^{\frac n k}_+,B_k+\sum_{i=1}^N q_i\p^{-1}r_i\right],\\
    &q_{i,t_n}=(B_k+\sum_{j=1}^N q_j\p^{-1}r_j)^{\frac n k}_+(q_i),\\
    &r_{i,t_n}=-(B_k+\sum_{j=1}^N q_j\p^{-1}r_j)^{\frac n k *}_+(r_i), \quad i=1,\cdots,N,
  \end{align}
\end{subequations}
which is the so-called $k$-constrained KP hierarchy \cite{MR1345081,MR1117170,MR1185854}.

For $k=2$ and $n=3$, (\ref{eqs:k-constrained}) gives rise to the
second type of KdV equation with self-consistent sources.
\begin{align*}
  &u_t=\frac14u_{xxx}+3uu_x+\frac34\sum_{i=1}^N(q_{i,xx}r_i-q_ir_{i,xx}),\\ 
  &q_{i,t}=q_{i,xxx}+3uq_{i,x}+\frac32 q_i\sum_{j=1}^Nq_jr_j+\frac32u_xq_i,\\
  &r_{i,t}=r_{i,xxx}+3ur_{i,x}-\frac32 r_i\sum_{j=1}^Nq_jr_j+\frac32u_xr_i,\quad i=1,\cdots,N.
\end{align*}
For $k=3$ and $n=2$, 
(\ref{eqs:k-constrained}) gives rise to the second
type of Boussinesq equation with self-consistent sources.
\begin{align*}
  &u_{tt}+\frac13u_{xxxx}+2(u^2)_{xx}-\frac43\sum_{i=1}^N (q_ir_i)_{xx}=0,\\
  &q_{i,t}=q_{i,xx}+2uq_i,\\
  &r_{i,t}=-r_{i,xx}-2ur_i,\quad i=1,\cdots,N.
\end{align*}

\section{Conclusions}
A method is proposed in this paper to construct a new extended KP hierarchy, which enables us to find the first and
second types of KPSCS (i.e., (\ref{eqns:KPSCS}) and (\ref{eqns:Another-KPSCS})) in a different way from those
in \cite{MR708435,MR910584,MR940618,MR2077716,Wang-HY} and to get their Lax representations directly. The new
extended KP hierarchy offers natural reductions to the well-known Gelfand-Dickey hierarchy with self-consistent
sources and to the $k$-constrained KP hierarchy. The $k$-constrained KP hierarchy includes the second type of
KdV equation with self-consistent sources (the Yajima-Oikawa equation) and the second type of Boussinesq
equation with self-consistent sources. The method proposed here provides a general way to construct the
soliton equations with self-consistent sources and their Lax representations. This approach for constructing
extended hierarchies can be applied to other $(2+1)$-dimensional systems (such as mKP ,BKP, CKP, etc) and
semi-discrete systems (e.g., 2-Toda hierarchy). We will present some other new extended hierarchies in the
forthcoming paper.

\section*{Acknowledgments}
This work was supported by National Basic Research Program of China (973 Program) (2007CB814800) and National
Natural Science Foundation of China (grand No. 10601028).


 \def\cprime{$'$} \def\cprime{$'$} \def\cprime{$'$} \def\cprime{$'$}
   \def\cprime{$'$}

\end{document}